\documentstyle[12pt]{article}       
\begin{document}
\newcommand{\ncd}{\newcommand} \newcommand{\rcd}{\renewcommand}
\ncd{\be}{\begin{equation}}        \ncd{\ee}{\end{equation}}
\ncd{\bq}{\begin{quote}}           \ncd{\eq}{\end{quote}}
\ncd{\barr}{\begin{array}}         \ncd{\earr}{\end{array}}
\ncd{\sectreset}[1]{\section{#1}\setcounter{equation}{0}}
\ncd{\sps}{{super\-selec\-tion}}   \ncd{\st}{{sta\-tistics}}
\ncd{\dmn}{{dimension}}            \ncd{\cjg}{{conjugation}}
\ncd{\obs}{{observable}}           \ncd{\sy}{{symmetry}}
\ncd{\cov}{{covariant}}            \ncd{\thy}{{theory}}
\ncd{\bgs}{{braid group \st}}      \ncd{\spt}{{space-time}}
\ncd{\pgs}{{permutation group \st}} \ncd{\spl}{{space-like}}
\ncd{\rep}{{repre\-sen\-tation}}   \ncd{\vrep}{{vacuum \rep}}
\ncd{\irrep}{{irreducible \rep}}   \ncd{\per}{{positive-energy \rep}}
\ncd{\emor}{{endo\-mor\-phism}}    \ncd{\amor}{{auto\-mor\-phism}}
\ncd{\hmor}{{homo\-mor\-phism}}    \ncd{\imor}{{iso\-mor\-phism}}
\ncd{\dmor}{{diffeo\-mor\-phism}}  \ncd{\lc}{{light-cone}}
\ncd{\ea}{{exchange algebra}}      \ncd{\ca}{{current algebra}}
\ncd{\itw}{{inter\-twiner}}        \ncd{\cf}{{con\-formal}}
\ncd{\og}{{ortho\-gonal}}          \ncd{\on}{{ortho\-normal}}
\ncd{\mb}{{M\"obius group}}        \rcd{\pc}{{Poincar\'e}}
\ncd{\rfb}{{reduced field bundle}} \ncd{\trafo}{{trans\-for\-ma\-tion}}
\ncd{\qft}{{quantum field \thy}}   \ncd{\emt}{{energy-momentum tensor}}
\ncd{\comrel}{{commutation relation}} \ncd{\typethree}{{type $I\!I\!I$}}
\ncd{\vn}{von~Neumann}      \ncd{\vna}{von~Neumann algebra}
\ncd{\cex}{conditional expectation}
\rcd{\aa}{{\cal A}}     \ncd{\ff}{{\cal F}}     \ncd{\oo}{{\cal O}}
\ncd{\bb}{{\cal B}}     \ncd{\hh}{{\cal H}}     \ncd{\kk}{{\cal K}}
\rcd{\a}{\alpha}  \rcd{\b}{\beta}  \ncd{\g}{\gamma}  \rcd{\d}{\delta}
\ncd{\po}{{\pi_0}}                 \ncd{\ovl}{\overline}
\ncd{\arr}{\! \rightarrow \!}      \ncd{\vphi}{\varphi}
\ncd{\sig}{{\sigma}}   \ncd{\eps}{\varepsilon}
\ncd{\NN}{{\sf I\!N}}  \ncd{\CC}{{}\;{\sf I\!\!\!C}}
\ncd{\ZZ}{{\sf Z\!\!Z}}
\ncd{\sub}[2]{#1 \subset #2}  \ncd{\comm}[1]{(#1)' \cap #1}
\ncd{\s}{S^1}    \rcd{\rho}{\varrho}
\hyphenation{mono-dromy}
\def\CMP#1{Com\-mun.\ Math.\ Phys.\ {\bf #1}}
\def\RMP#1{Rev.\ Math.\ Phys.\ {\bf #1}}
\def\NP#1{Nucl.\ Phys.\ {\bf #1}}
\def\PL#1{Phys.\ Lett.\ {\bf #1}}
\def\PR#1{Phys.\ Rev.\ {\bf #1}}
\def\JFA#1{J.\ Funct.\ Anal.\ {\bf #1}}
\def\etal{{\it et al.}}
\def\KHR{K.-H.\ Rehren} \def\BS{B.\ Schroer} \def\KF{K.\ Fredenhagen}
\def\JF{J.\ Fr\"ohlich} \def\DB{D.\ Buchholz} \def\RH{R.\ Haag}
\def\SD{S.\ Doplicher} \def\JR{J.E.\ Roberts} \def\RL{R.\ Longo}
\parindent0mm \parskip2mm \rcd{\thefootnote}{\fnsymbol{footnote}}
\rightline{DESY 93-116}
{\Large \bf Subfactors and Coset Models\footnote{Talk at the Workshop on
``Generalized Symmetries in Physics'', Clausthal (FRG), July 1993, \\
to appear in the proceedings, eds.\ H.-D.\ Doebner \etal}}
\\[9mm]
{\large Karl-Henning Rehren }\\[3mm]
II.\ Inst.\ Theor.\ Physik, Univ.\ Hamburg, D-22761 Hamburg (FR Germany)
\\ {\small e-mail: i02reh@dsyibm.desy.de} \\[5mm]
\rcd{\thefootnote}{\arabic{footnote}} \addtolength{\baselineskip}{-1pt}
\bq
 {\bf Abstract:} Some facts about von Neumann algebras and finite
 index inclusions of factors are viewed in the context of local \qft.
 The possibility of local fields intertwining \sps\ sectors with \bgs\
 is explored. Conformal embeddings and coset models serve as examples.
 The associated \sy\ concept is pointed out.
\eq
\addtolength{\baselineskip}{1.5pt}
\section{Introduction}
\parskip3mm \rcd{\theequation}{\thesection.\arabic{equation}}
The present article is a pedestrian's view of some basic mathematical
facts about \vna s and the structure theory of subfactors relevant for
the physical problem of implementability of \sps\ sectors by charged
fields. Its intention is to facilitate the access to abstract results by
formulating them in terms of algebraic relations of rather direct
physical significance, and to illustrate them by a large class of models
(coset models of \cf\ \qft) well known to a broad audience. The present
work is a precursory study related to a joint program in progress with
\KF, \RL, and \JR\ about local extensions of local quantum field
theories.

The general analysis we shall present below was incited by a recent
result of a model study of chiral \ca s \cite{Bul}, which at first sight
quite puzzled the author. The authors of \cite{Bul} were searching for
{\it local} extensions of chiral $SU(2)$ \ca s which are identified by
their local conformal block functions satisfying the $SU(2)$ Ward
identities. They found a solution at level 10 for the isospin 3 sector,
clearly different from the standard non-local vertex operator (Coulomb
gas) solution.

The ``fusion rules'' which are read off the local solution are
\be [3] [3] = [0] + [3] \ee
in contrast to the standard fusion rules
\be [3] [3] = [0] + [1] + [2] + [3] + [4] .\ee
This seems to contradict the message from the general theory of \sps\
sectors \cite{DHR} that the fusion rules are {\it intrinsic} to a given
local \qft. Moreover, the isospin 3 sector is known to have non-trivial
\bgs, so there is the surprising fact that one can associate it with
local correlation functions. In fact, the local solution found in
\cite{Bul} has been identified as the conformal embedding of the $SU(2)$
\ca\ at level 10 into the $SO(5)$ (or $Sp(4)$) \ca\ at level 1. It is
well known that the \vrep\ of the latter splits into the \rep s $[0]$
and $[3]$ of the former, where the primary field for the \rep\ $[3]$ is
the isospin 3 multiplet of $SO(5)$ currents orthogonal to the embedded
$SU(2)$ currents. The \cf\ blocks and the fusion rules (1.1) are just
the correlation functions and the operator product expansions of these
$SO(5)$ currents.

The second solution found in \cite{Bul} is much less surprising. It is
the \cf\ embedding \cite{BBSW} of $SU(2)$ at level 4 into $SU(3)$ at
level 1, generating the isospin 2 sector. This sector has \pgs, and the
fusion rule $[2][2] = [0]$ coincides with the standard one. In fact, the
detailed analysis shows that the $SU(2)_4$ theory is exactly the
subtheory of $\ZZ_2$-invariants of the $SU(3)_1$ theory, where the
global \sy\ group $\ZZ_2$ is the involutive Lie group \amor\ of $SU(3)$
associated with the symmetric space $SU(3)/SU(2)$, lifted to the \ca.

The previous solution is an example of a model, where a sector of a
local \qft, which has genuine \bgs, is generated by local fields of some
larger local \qft. In fact, \cf\ coset models generically exhibit this
feature, and the $SU(2)_4 \subset SU(3)_1$ case is rather the exception.
The aim of the present article is a general analysis of this scheme, as
an alternative to the widespread conviction that charged fields
generating sectors with \bgs\ should exhibit braid-like \comrel s
at \spl\ distance.

The apparent conflict between locality and \bgs\ dissolves due to some
intrinsic ``defect''. As we shall see, one can only find {\it
intertwining} fields
\be \psi a = \rho(a) \psi \ee
for the \emor\ corresponding to the sector in question \cite{DHR},
but not {\it implement} the \emor\ itself:
\be \rho(a) \stackrel!= \sum_i \psi^i \, a \, \psi^{i*} \ee
as it is the case with \pgs\ \cite{DR2}. The defect is responsible for
a mismatch between the abstract fusion rules of the theory of \sps\
sectors \cite{DHR} and the multiplicative structure (operator product
expansions) of the generating fields. This then solves the previously
mentioned puzzle. Namely, the fusion rules (1.1) are not the intrinsic
ones but refer to the embedding into a larger theory.

In order to place the above example in a general context, and to
elucidate the point where the standard reasoning fails, we resort to the
theory of \vna s and subfactors. By the comparison of the general case
of a subfactor $\sub AB$ with the case when the subfactor is given by
the invariants under the action of a \sy\ group, we shall very
explicitly see the role of the gauge group (of the first kind) to
exclude the occurrence of the defect, thus leading to the standard
duality theory \cite{DR2}.

The general case is described by a faithful normal \cex\
\be \mu : B \arr A \ee
i.e.\ a positive unit preserving map of $B$ onto $A$ satisfying
\be \mu(abc) = a\mu(b)c \qquad (a,c \in A,\; b \in B). \ee
A \cex\ generalizes the abstract properties of a group average
\be \mu(b) := |G|^{-1}\sum_{g \in G} \a_g(b), \ee
which special case we shall always compare to the general situation.

Associated with the \cex\ is a ``canonical'' \emor\ $\g : B \arr A$, the
structure theory of which is treated in Sect.\ 2. It is of direct
physical significance: the sector decomposition of $\gamma$ as an \emor\
of $B$ allows to recover the \sy\ group in the case (1.7), and naturally
leads to Ocneanu's generalized \sy\ concept (``paragroups'' \cite{Oc})
in the general case, see Sect.\ 4. The sector decomposition of $\gamma$
as an \emor\ of $A$ corresponds to the decomposition of a \rep\ of $B$
upon restriction to $A$, provided the former is obtained from an
invariant state $\omega_0 = \omega_0 \circ \mu$. In the quantum field
theoretical context (i.e.\ $A$ and $B$ are a pair of local \vna s from
local quantum field theories $\aa$ and $\bb$ and $\omega_0$ is the
vacuum state in which the \sy\ is unbroken), this is the ``branching
rule'' for the vacuum sector of the larger theory. The sectors of the
subtheory thus obtained are interpolated by fields from the larger
theory.

The canonical \emor\ uniquely characterizes the underlying inclusion
$\sub AB$. We show how to recover detailed structure information about
the latter in terms of intertwining operators and their algebraic
properties. In Haag's spirit that ``fields are just coordinates on local
algebras'', we are putting coordinates on an inclusion.

Our conclusion in the quantum field theoretical context is that indeed
some sectors with \bgs\ can be generated by a local field algebra. The
exact relation to the \st\ is discussed in Sect.\ 3. In fact, this
possibility is realized in (almost) all coset models of \cf\ \qft.

The models we have in mind include the previously mentioned \cf\
embedding $SU(2)_{10} \subset SO(5)_1$, but also coset models \cite{GKO}
of the form
\be Diag(G \times G)_{k+l} \otimes W \subset G_k \otimes G_l \ee
where $G_k$ stands for the level $k$ chiral \ca\ of the compact
simple Lie group $G$, or still more generally
\be H_l \otimes W \subset G_k \ee
where $\sub HG$ is a pair of compact semi-simple Lie groups (not
necessarily simple such that the levels may have several components for
the simple components of $H$ and $G$). In all cases, $W$ stands for a
local \qft\ decoupled from the subgroup currents. It is trivial for \cf\
embeddings and otherwise contains the coset \emt\ given by
\be T^{(G)} = T^{(H)} + T^{\rm coset} \ee
in terms of the Sugawara expressions for $T^{(G)}$, $T^{(H)}$. If the
central charge of $T^{\rm coset}$ is $c>1$, then $W$ may contain further
primary fields of higher scaling \dmn s.

We consider a coset model as an extension of the local theory of \obs s
$\aa$ into a local theory of ``charged'' fields $\bb$. The detailed
knowledge of these models precisely confirms and exemplifies our general
findings. In Sect.\ 4, we briefly sketch the generalized \sy\ associated
with such an extension.

According to Wassermann \cite{Wa}, the local \vna s of \ca\ models are
defined -- in purely group theoretical terms -- as $\pi_0(L_IG)''$ where
$LG$ is the group of smooth maps (loops) $: S^1 \arr G$, the local
subgroup $L_IG$ consists of those maps with support in the interval $I$
of the circle, and $\pi_0$ is the projective \vrep\ of level $k$.
The term ``vacuum'' refers to the implementation of the group of \dmor s
of the circle which act as \amor s of the loop group, and in particular
to the spectrum of the ``rigid'' rotations, while the level refers to
the cocycle associated with a projective \rep. Similar definitions can
be given for the algebra of the (coset) \emt\ in terms of the \dmor\
group of $\s$ and its local subgroups.

Following \cite{Wa}, we shall argue in Sect.\ 5 that the index of the
inclusion of the corresponding local \vna s can be finite only when the
\vrep\ of $\bb$ decomposes finitely into \rep s of $\aa$, and give
a formula to compute the index. The branching rules for coset models
being known, this formula is very explicit.

The contents of this article are not really new, although we sometimes
adopt an unconventional point of view. By viewing a class of physical
models in the light of the abstract structure theory for the canonical
\emor, some model findings are systematically related to the general
algebraic structures which underly local extensions of local quantum
field theories. Moreover, the latter are formulated in physically
evident terms. Some novel insight concerns the properties of
intertwining fields and the relation between their local commutativity
and \bgs, the precise identification of the points where more general
subtheories depart from the rather special gauge \sy\ inclusions, as
well as some algebraic control over the defect going along with
non-integer statistical \dmn s.

\sectreset{Inclusions of type {\it \bf III} factors}
We collect some well known results about finite index inclusions of
\typethree\ factors \cite{PP,Ko,Hi,Lo2,Lo3}, which we shall need for the
subsequent discussion, and qualify the abstract statements in the
special case when the inclusion is given by the action of a \sy\ group.
The generalization to Hopf algebra actions is also known \cite{Lo4}.

We consider an irreducible inclusion $\sub AB$ of \typethree\ factors.
With this pair, we have in mind a local \vna\ $\bb(\oo)$ from some \qft\
and a local \vna\ $\aa(\oo)$ from some subtheory. However, the purely
mathematical statements in this section do not refer to the physical
context. The local aspects will be treated in Sect.\ 3.
\bq
 {\bf Proposition 1:} Let $\sub AB$ be an irreducible inclusion of
 \typethree\ factors with finite index $\lambda$. There are a unique
 normal \cex\ $\mu: B \arr A$ which satisfies the operator estimate
 \be \mu(bb^*) \geq \lambda^{-1} \cdot bb^* \qquad (b \in B), \ee
 and an isometry $V \in B$ such that
 \be \barr{lc} (i) & \mu(bV^*)V = V^*\mu(Vb) = \lambda^{-1}b \qquad
 (b \in B) \\ (ii) & \mu(VV^*) = \lambda^{-1}. \earr \ee
 The map $\g : B\arr \sub AB$
 \be \g(b) := \lambda \cdot \mu(VbV^*) \ee
 is an \emor, called the canonical \emor, and $V$ is an \itw\ $: id
 \arr \g$, i.e.
 \be V \, b = \g(b) \, V. \ee
\eq
The canonical \emor\ was originally introduced in \cite{Lo1} in terms of
modular conjugations. It is very useful in \rep\ theory, since it allows
the restriction of an \emor\ $\sigma$ of $B$ to an \emor\ $\g \circ
\sigma\vert_A$ of $A$, and the induction of an \emor\ $\sigma$ of $A$ to
an \emor\ $\sigma\circ\gamma$ of $B$. Considering \emor s as \rep s of
spatial \vna s, these prescriptions correspond to the restriction and
(Mackey) induction of \rep s.

We specify the statement of Prop.\ 1 in the case when $A$ are the
``gauge invariants'' of $B$ with respect to some \sy\ group.
\bq
 {\bf Proposition $\bf 1'$:} Let in Prop.\ 1 $A = B^\a$ be the fixpoint
 subalgebra under an outer action $\a : G \arr Aut(B)$ of a \sy\
 group $G$. Then $\lambda = |G|$ (in particular, $G$ is finite), and
 $\mu$ is the group average:
 \be \mu = |G|^{-1} \sum_{g \in G} \a_g\,. \ee
 The operators
 \be V_g := \a_g(V) \ee
 form a complete system of \on\ isometries:
 \be V_g^*V_h = \delta_{gh} \qquad {\rm and} \qquad
 \sum_{g \in G} V_gV_g^* = 1, \ee
 carrying the left regular \rep\ of $G$:
 \be \a_g(V_h) = V_{gh}. \ee
 The canonical \emor\ is
 \be \g(b) = \sum_{g \in G} V_g \a_g(b) V_g^*. \ee
\eq
The completeness of $V_g$ ist just (2.2($ii$)): $|G| \cdot \mu(VV^*) =
1$. The \on ity is verified by summing $V^*\a_g(VV^*)V$ over the group,
which implies $\sum_{g \neq e} (V^*V_g)(V^*V_g)^* = 0$ and therefore
$V^*V_g = \delta_{ge}$. (2.9) follows from the definitions.

Next, we discuss the reducibility of \emor s. Let $\sigma : M
\arr M$ be an \emor\ of a \typethree\ factor $M$. It is reducible iff
the relative commutant $\sigma\comm M$ is non-trivial. Every projection
$e \in \sigma\comm M$ corresponds to a sub-\emor\ $\sigma_e$ defined as
follows. Pick an isometry $w \in M$ such that $ww^* = e$. Then
\[ \sigma_e(\cdot) := w^* \sigma(\cdot) w. \]
By construction, $w$ is an \itw\ $: \sigma_e \arr \sigma$. If
\[ 1 = \sum_s e_s \qquad {\rm and} \qquad  e_s = \sum_i e_s^i \]
are the central partition of unity in $\sigma\comm M$, and partitions of
$e_s$ into minimal projections in $\sigma\comm M$, respectively, then
the corresponding sub-\emor s $\sigma_s^i = w_s^{i*} \sigma(\cdot)
w_s^i$ fall into inner equivalence classes (sectors) labelled by $s$,
while $i = 1,\ldots n_s$ counts the multiplicity of the sector $[s]$
within $\sigma$. It is possible and convenient to choose $\sigma_s^i =
\sigma_s$ independent of $i$, and to write
\[ \sigma \simeq \bigoplus_s n_s \; \sigma_s. \]
For reducible inclusions one defines $Index(\sub NM)$ to be the
{\it minimal}\ index \cite{Hi}, and calls $d(\sigma) :=
\sqrt{Index[\sub {\sigma(M)}M]}$ the {\it \dmn} of an \emor. The \dmn\
is additive and multiplicative:
\[ d(\sigma) = \sum_s n_s d(\sigma_s) \qquad {\rm and} \qquad
d(\sigma\rho) = d(\sigma) d(\rho).\]
\bq
 {\bf Proposition 2:} Let $\sub AB$ be as in Prop.\ 1. Denote by
 $\rho : A \arr A$ the restriction of $\g$ to $A$. Then
 \be \rho(A) \subset \g(B) \subset A \subset B \ee
 is (the beginning of) a Jones tower. In particular, all inclusions are
 irreducible with index $\lambda$ and $d(\g) = d(\rho) = \lambda$. Both
 $\sub {\g(B)}B$ and $\sub {\rho(A)}A$ are reducible (unless $\lambda =
 1$):
 \be (a) \quad \rho \simeq \bigoplus_s N_s \; \rho_s \qquad {\rm and}
 \qquad (b) \quad \g \simeq \bigoplus_t M_t \; \g_t \,, \ee
 either decomposition containing the identity as a sub-\emor\ with
 multiplicity 1.

 Let $W \in A$ be the \itw s $: \rho_s \arr \rho$ according to the above
 decomposition theory for $\rho$, and $V \in B$ as in Prop.\ 1. The map
 $W \mapsto \psi := W^*V$ is bijective onto the linear space $\hh_s
 \subset B$ of operators satisfying
 \be \psi a = \rho_s(a) \psi \qquad (a \in A). \ee
 $\hh_s$ is an $N_s$-\dmn al Hilbert space of isometries with scalar
 product $\psi^*\psi' \in A' \cap B = \CC$.
\eq
In the quantum field theoretical context, $A$ are local \obs s
$\aa(\oo)$, $B$ are local fields $\bb(\oo)$, and $\rho_s$ are
(localized) \emor s of $\aa$ representing the \sps\ sectors \cite{DHR},
see Sect.\ 3. We have thus established the existence of intertwining
fields for all sectors contained in $\rho$, and conversely all sectors
which have intertwining fields in $\bb$ are contained in $\rho$.

{\it Proof of the last statements of Prop.\ 2}: The map $W \mapsto
\psi$ is injective since $\mu(\psi\psi^*) = \lambda^{-1} W^*W > 0$. It
is inverted by $\psi \mapsto W:= \lambda \mu(V\psi^*) \in A$. The stated
intertwining properties of $\psi$ resp.\ $W$ are readily checked.

The following specification of Prop.\ 2 in the group symmetric case is
easily obtained with group theoretical arguments.
\bq
 {\bf Proposition $\bf 2'$:} Let $A = B^\a$ be as in Prop.\ $1'$.
 Then
 \be (a) \quad \rho \simeq \bigoplus_r dim(r) \rho_r \qquad {\rm and}
 \qquad (b) \quad \g \simeq \bigoplus_{g \in G} \a_g \ee
 where in $(a)$ the sum extends over the unitary \irrep s
 $\tau^r$ (of \dmn\ $dim(r)$) of the \sy\ group $G$. For every sector
 $[r]$ and an \on\ basis of \itw s $W_r^i : \rho_r \arr \rho$ in $A$,
 the corresponding charged fields $\psi^i = \sqrt{|G|/dim(r)} \cdot
 (W_r^i)^*V \in \hh_r \subset B$ are a complete set of \on\ isometries:
 \be (i) \quad \psi^{i*}\psi^j = \delta_{ij} \qquad \quad
 (ii) \quad \sum_{i=1}^{dim(r)} \psi^i\psi^{i*} = 1 \ee
 which transform linearly in the \rep\ $\tau^r$:
 \be \a_g(\psi^i) = \sum_{j=1}^{dim(r)} \psi^j\cdot\tau^r_{ji}(g) \ee
 and implement the sectors:
 \be \rho_r(a) = \sum_{i=1}^{dim(r)} \psi^i \, a \, \psi^{i*}. \ee
\eq
The reader will find the sketch of a constructive proof in the appendix.

With Prop.\ $1'$ we have recovered the Cuntz algebra (= $C^*$ algebra
generated by a complete set of \on\ isometries) of intertwining fields
which implement the sectors. In the context of \sps\ sectors in local
\qft, it was first derived in \cite{DR1}, and is the basic tool for the
reconstruction \cite{DR2} of the gauge symmetric field algebra
from the sector structure of the \obs s.

If one looks through the argument for the completeness relation
(2.14($ii$)) (see the appendix) one finds that it owes its specific
form to the linear transformation law (2.15). Namely, by our general
construction, for every $\psi \in\hh_s$ the ``average'' of $\psi\psi^*$
is a multiple of unity:
\be \mu(\psi\psi^*) = \lambda^{-1} \cdot W^*W \propto 1 .\ee
On the other hand, introduce the joint range projection of the Hilbert
space of isometries $\hh_s$ in terms of any \on\ basis $\psi^i \in
\hh_s$
\be E_s := \sum_{i=1}^{N_s} \psi^i \psi^{i*}.\ee
Now, under the {\it linear} group action (2.15), the average
$\mu(\psi\psi^*)$ is a multiple of $\sum_i \psi^i\psi^{i*}$, hence
$E_s = 1$. In the general case, however, (2.17) is just an abstract
relation in $B$ rather than an algebraic relation among the operators
$\psi \in \hh_s$, while $E_s$ will be a non-trivial projection in
$\rho_s(A)' \cap B$. This ``defect'' $E_s < 1$ is related to the
mismatch between the multiplicity $N_s$ and the dimension $d(\rho_s)$
in (2.11(a)):
\bq
 {\bf Proposition 3:} If $\sub AB$ has finite index, then the
 multiplicities $N_s$ in (2.11($a$)) are not larger than the \dmn s
 $d(\rho_s)$:
 \be N_s \leq d(\rho_s). \ee
 Equality holds iff the defect projection $E_s$ is unity.
\eq
{\it Proof:} For $\psi^i \in \hh_s$ as above, the operators $X_i :=
\g(\psi^i)$ are a system of orthogonal \itw s in $A$
\[  X_i \rho(a) = \rho\rho_s(a) X_i\,. \]
Hence the $N_s$ projections $X_iX_i^* \in \rho\rho_s\comm A$ correspond
to $N_s$ disjoint sub-endo\-mor\-phisms equivalent to $\rho$ contained
in $\rho\rho_s$. The additivity and multiplicativity of \dmn s implies
\[ N_s \cdot d(\rho) \leq d(\rho\rho_s) = d(\rho)d(\rho_s) \]
where equality holds iff $\sum_i X_iX_i^*$ is a partition of unity.
Since $d(\rho) = \lambda$ is finite, the inequality (2.19) follows,
and since $\sum_i X_iX_i^* = \g(E_s)$, equality holds iff $E_s = 1$.

In the group \sy\ case, we conclude $d(\rho_r) = N_r = dim(r)$ and
$\rho\rho_s \simeq d(\rho_s) \cdot \rho$. The \dmn\ formula for finite
groups, $|G| = \sum_r dim(r)^2$, is recovered from (2.13). It becomes
also clear from Prop.\ 3 that sectors with non-integer \dmn\
$d(\rho_s)$ must suffer the defect. Finally, the defect is related to
the depth \cite{Oc} of the inclusion $\sub AB$: suppose the defect to be
absent in all sectors $\rho_s$. Then together with $\rho_s$ all
sub-\emor s of $\rho_s\rho_r$ are implemented, and by Prop.\ 2 are
already contained in $\rho$. Then $\g(B) \subset A$ and $\sub AB$ have
depth 2 and $\sub AB$ is given by the action of a Hopf algebra
\cite{Lo4}.

The failure of Prop.\ 3 at infinite index will be exemplified in Sect.\
6.

\sectreset{The relation to statistics}
We have so far considered a single pair of \vna s $\sub AB$. A local
\qft\ is given by a local net, i.e.\ an assignment of an algebra of
local \obs s $\aa(\oo)$ to every \spt\ region $\oo$ such that local
algebras at \spl\ distance commute. These algebras are \typethree\
factors for typical bounded (``double cone'') regions \cite{BDF}. In the
case of chiral \ca s, the local algebras $\aa_I$ are assigned to the
intervals $I$ of the circle. $\aa_I$ are also \typethree\ factors
\cite{Wa,FG}. The total algebra $\aa$ of \obs s is the $C^*$ algebra
generated by all its local subalgebras.

A theory is covariant with respect to the \spt\ \sy\  group (the \pc\
group or, on the circle, the \mb) if the latter acts by \amor s
on the local net, i.e.\
\be \a_x(\aa(\oo)) = \aa(x\oo) .\ee
A \rep\ is covariant if the \amor s $\a_x$ are implemented by unitary
operators in the Hilbert space. The spectrum condition refers to the
positivity of the generators of the translations in a given \rep.

Let us now turn to a subtheory. By this we mean a pair of local nets
$\aa$ and $\bb$ such that for every region
\be \aa(\oo) \subset \bb(\oo), \ee
and the covariance \amor s $\a_x^{(\bb)}$ of $\bb$ coincide on $\aa$
with $\a_x^{(\aa)}$ (we shall thus drop the distinction).

The analysis of Sect.\ 2 then applies to every single inclusion (3.2).
It is important to note that the canonical \emor\ $\g$ certainly, and
the \cex\ $\mu$ possibly depends on $\oo$. There arise thus consistency
problems when extending the maps $\mu, \g, \rho$ globally. These
problems are not subject of the present article. Let us therefore --
rather than deriving it from some more general principles -- make the
certainly very reasonable assumption that $\mu$ is consistently defined
for all local algebras (i.e.\ $\mu^{(\oo_1)}\vert_{\bb(\oo_2)} =
\mu^{(\oo_2)}$ if $\oo_2 \subset \oo_1$) and commutes with $\a_x$.
Furthermore, we shall assume that the vacuum state $\omega_0$ of $\bb$
is the unique \pc\ invariant state. Then $\omega_0$ is also invariant
under the \cex:
\be \omega_0 \circ \a_x = \omega_0 = \omega_0 \circ \mu \;. \ee
We fix $\oo_0$ and apply Prop.\ 1 and 2 to the inclusion $\aa(\oo_0)
\subset \bb(\oo_0)$. In particular, $V$ is an isometry in $\bb(\oo_0)$.
Then we extend $\gamma$ globally by (2.3):
\be \g(b) := \lambda\cdot\mu(VbV^*) \qquad\quad (b \in \bb). \ee
By locality, (1.6), and (2.2), $\rho := \g\vert_\aa$ is an \emor\ of
$\aa$ localized in $\oo_0$, namely it acts trivially on $a \in
\aa(\oo')$ if $\oo'$ is \spl\ separated from $\oo_0$. The same is not
true for $\g \in End(\bb)$, nor does $\g$ map $\bb(\oo)$ into
$\aa(\oo)$ unless $\oo$ contains $\oo_0$.

The operators $\psi \in \hh_s \subset \bb(\oo_0)$ are local \itw s for
the sub-\emor s $\rho_s$ of $\rho$ which are also localized in $\oo_0$.

Next, let $\rho$ correspond to a covariant \rep\ of $\aa$. Then
\cite{DHR} there is a unitary cocycle $x \mapsto U_x \in \aa$
\be U_{xy} = \a_x(U_y)U_x \ee
such that $\rho_x := Ad_{U_x} \circ \rho = \a_x \rho \a_x^{-1}$ is
equivalent to $\rho$ and localized in $x\oo_0$. Then also the sub-\emor
s $\rho_s$ are covariant (with cocycle $U^s_x$). We claim:
\bq
 {\bf Lemma:} \hfill $\a_x(\psi) = U^s_x \cdot \psi \qquad
 (\psi \in \hh_s)$. \hfill (3.6)
\eq \setcounter{equation}{6}
{\it Proof:} It is easily checked that $(U^s_x)^*\a_x(\psi)$ are again
\itw s for $\rho_s$ and therefore belong to $\hh_s$. Thus the map
$(x,\psi) \mapsto (U^s_x)^*\a_x(\psi)$ is a finite \dmn al unitary
\rep\ of the \pc\ or \mb\ on $\hh_s$. Since every such \rep\ is trivial,
one gets (3.6).

Locality of $\bb$ implies that $\a_x(\psi)$ and $\psi$ commute when
$x\oo_0$ is \spl\ to $\oo_0$. Therefore ($i,j = 1,\ldots N_s$)
\[ U^s_x\psi^i\psi^j = \psi^jU^s_x\psi^i = \rho_s(U^s_x)\psi^j\psi^i .\]
Now, $U_x^{s*}\rho_s(U^s_x)$ is the \st\ operator $\eps_s =
\eps(\rho_s,\rho_s)$ \cite{DHR}, so locality reads
\be \eps_s\,\psi^i\psi^j = \psi^j\psi^i .\ee
When there is no defect as discussed in Prop.\ 3, this can be solved for
$\eps_s$:
\be  \eps_s = \sum_{ij} \psi^j\psi^i\psi^{j*}\psi^{i*} \ee
implying $\eps_s^2 = 1$, i.e.\ \pgs\ as expected for sectors
implemented by local fields. In contrast, when there is a defect, one
only gets
\be \eps_s^2\psi^i\psi^j = \psi^i\psi^j ,\ee
i.e.\ the non-trivial joint range projection of $\psi^i\psi^j$ is an
eigenprojection of the monodromy operator with eigenvalue 1. On the
other hand, the monodromy operator is diagonalized by the projections
for the decomposition of $\rho_s^2 \simeq \bigoplus \rho_k$ into its
irreducible components \cite{DHR}. Thus, although $\psi^i\psi^j$ are
\itw s for $\rho_s^2$, they do not project to \itw s for irreducible
sub-\emor s $\rho_k$ unless the monodromy eigenvalue
\be \frac{\kappa(\rho_k)}{\kappa(\rho_s)^2} \stackrel != 1. \ee
Here $\kappa$ is the \st\ phase \cite{DHR} related to the (fractional)
spin of a sector. Namely, every candidate \itw\ $T^*\psi^i\psi^j \in
\bb$ (with $\aa \ni T : \rho_k \arr \rho_s^2$) for a subsector $\rho_k$
violating (3.10) must vanish. This immediately explains the
``truncation'' of the fusion rules (1.1) as compared to the intrinsic
fusion rules (1.2).

Actually, one can prove more, thereby restricting the branching
rules for the \vrep\ in the first place. (3.10) is then a
trivial consequence.
\bq
 {\bf Proposition 4:} All subsectors of $\rho$ have \st\ phase
 $\kappa(\rho_s) = 1$.
\eq
{\it Proof:} The ``master'' equation underlying (3.7):
\be \eps_\rho VV = VV, \ee
where $\eps_\rho = \eps(\rho,\rho)$, obtains in the same way as (3.7).
Introduce isometric \itw s in $A$: $W := \lambda^{1/2}\mu(V) : id \arr
\rho$ and $R := \g(V)W : id \arr \rho^2$, and observe $V^*R =
\lambda^{-1/2}V$. Define the left-inverse $\phi_\rho(a) := R^*\rho(a)R$
and compute $d(\rho)\phi_\rho(\eps_\rho) = \lambda\mu(V^*\eps_\rho V)$.
Next, compute ($\hat R \equiv \lambda^{1/2}R$)
\[ V^*\eps_\rho V = \hat R^*V\eps_\rho V = \hat R^*\rho(\eps_\rho)VV =
\rho(\hat R^*)\eps_\rho^* VV = \rho(\hat R^*)VV = V \hat R^*V = VV^*. \]
The equality in the middle comes from the theory of \st\ \cite{DHR}.
Therefore,
\be \kk_\rho := d(\rho)\phi_\rho(\eps_\rho) = 1. \ee
Thus $\phi_\rho$ is the standard left-inverse of $\rho$, and since the
spectrum of $\kk_\rho$ is given by the \st\ phases $\kappa(\rho_s)$ of
all subsectors of $\rho$ \cite{DHR}, the claim follows.

The lemma and the constraint (3.10) derived from it, as well as
Prop.\ 4 fail if $\sub AB$ has infinite index (see Sect.\ 6).

\sectreset{The generalized \sy}
An alternative way to show the completeness (2.14($ii$)) of the
charged isometries in the group symmetric case relies on the gauge
principle: since the defect projection $E_s \in \rho_s(A)' \cap B$ is
gauge invariant by the linear transformation law (2.15), it is actually
contained in $A$ and therefore must be a scalar.

In the general case, an explicit formula like (2.15) is lacking. The
gauge transformations $\a_g$ -- which by (2.13($b$)) are the irreducible
components in the decomposition of the canonical \emor\ -- are replaced
by the \emor s $\g_t$ (2.11($b$)). The latter will not act linearly on
the \itw\ spaces $\hh_s$ of charged field operators. Yet, it seems
natural to consider the action of $\g_t$ on $B$ as the generalized \sy.

The questions arise what sort of a generalized group $\g_t$ form, and
in which sense $A$ consists of invariant quantities. We shall not
elaborate on this program here, but only include three remarks to
illustrate our conception of (gauge) \sy: everything that is apt to
characterize the position of a subalgebra in an algebra (the gauge
invariant quantities among all fields) is a good candidate for a
generalized symmetry.

1.) The underlying structure to $\g_t$ and all sub-\emor s of their
products is described by the even part of the bipartite graph associated
with the inclusion $\sub AB$ as discussed by Ocneanu \cite{Oc}. The odd
part of the graph is obtained if one writes the canonical \emor\ in the
form
\be \g = \sigma\bar\sigma  \qquad {\rm where} \quad \sigma(B) = A. \ee
The odd part of the graph then comprises all subsectors of $\gamma^n
\sigma$. In (4.1), $\sigma$ is an irreducible \emor\ of $B$ (since $\sub
AB$ is irreducible), and $\bar\sigma$ is a conjugate \emor\ uniquely
determined up to inner equivalence by the requirements that $\bar\sigma$
is irreducible and $\sigma\bar\sigma$ contains the identity.

2.) If it is only known that all irreducible sub-\emor s $\g_t$ of $\g$
in (2.11($b$)) have \dmn\ 1, i.e.\ are \amor s, then one can show
\cite{Lo3,KHR} that the latter may be (uniquely) chosen within their
inner equivalence classes such that they form a group of order $\lambda$
pointwise preserving $A$. In other words, it is possible to recover the
\sy\ group $G$ of $A = B^\a$ from its canonical \emor\ $\g : B \arr A$.
The argument is simple: since for every $t$, $\g_t^{-1}\g = \g_t^{-1}
\sigma \circ \bar\sigma$ again contains the identity, one concludes
that $\g_t^{-1}\sigma$ is conjugate to $\bar\sigma$ and therefore
equivalent to $\sigma$, and so is $\g_t\sigma$. It follows that
$\g_t^{-1}\g$ and $\g_t\g$ contain the same sectors as $\g$, so the
sector multiplication of $\g_t$ must form a group. Now, if $\g_t\sigma
= Ad_{U_t}\circ \sigma$, then $\a_t := Ad_{U_t^*}\circ \g_t$ preserve $A
= \sigma(B)$ pointwise and satisfy the composition law of the group up
to inner conjugation with some cocycle in $B$. Since the cocycle
commutes with $A$ it is scalar and the inner conjugation is trivial.

3.) There is a generalized ``gauge principle'' which allows to recover
(up to inner conjugation) the observables $A$ from the knowledge
of the \emor s $\g_t \in End(B)$ together with their multiplicities.
Namely, according to (2.11($b$)) one may construct $\g \in End(B)$ which
has the abstract properties characterizing canonical \emor s. E.g.,
there is a pair of isometric \itw s $V : id \arr \g$ and $W : \g \arr
\g^2$ satisfying the identities \cite{Lo4}
\be \barr{c} V^*W = \lambda^{-1/2} = \g(V^*)W \\ \g(W)W = WW, \qquad
\g(W^*)W = WW^*. \earr \ee
 From this one recovers the \cex\ $\mu(\cdot) := W^*\g(\cdot)W$ and the
observables $A := \mu(B) =$ fixpoints of $\mu$. Conversely, $\mu, V,
\g$ are the data of Prop.\ 1 associated with $\sub AB$, and $W =
\lambda^{1/2}\mu(V)$.

These remarks justify our proposal for the generalized \sy\ underlying
the inclusion $\sub AB$.

\sectreset{The index of coset models}
Let us now turn to the specific coset models of \cf\ \qft. It is known
\cite{Wa} that for \ca s as well as for the algebras of the \emt\ on the
circle, Haag duality holds in the \vrep, i.e.\ the local \vna s
associated with complementary intervals $I$ and $I^c = \s \setminus I$
are each other's commutants:
\be \po(\aa_I) = \po(\aa_{I^c})'. \ee
In other \rep s, Haag duality will be violated in general. Consider
the inclusion (locality!)
\be \pi(\aa_I) \subset \pi(\aa_{I^c})'. \ee
If $\pi = \po \circ\rho$ is given by an \emor\ of $\aa$ which is
localized in $I$, i.e.\ $\rho(a') = a'$ for $a' \in \aa_{I^c}$,
then due to (5.1), (5.2) turns into
\be \rho(\aa_I) \subset \aa_I\, . \ee
The index of the inclusions (5.2), (5.3) is given by the square of the
statistical \dmn\ $d(\pi)^2 \equiv d(\rho)^2$ of the \emor\ $\rho$
\cite{Lo2}, i.e., the notion of statistical \dmn\ for localized
\emor s of the $C^*$ algebra $\aa$ and the notion
                                       of \dmn\ for \emor s of a local
\vna\ $\aa_I$ as introduced in Sect.\ 2 coincide.

Now, denote the pairs of algebras of the coset models (1.8), (1.9)
by $\sub \aa\bb$. Clearly, the inclusion holds also for the local
algebras: $\aa_I \subset \bb_I$. Let $\pi$ be a \per\ of $\bb$ and
$\pi_|$ the restriction of $\pi$ to $\aa$. The following idea is due
to Wassermann \cite{Wa}. Consider the chain of inclusions
\be \pi_|(\aa_I)\;\stackrel\a\subset\; \pi(\bb_I) \;\stackrel\b\subset\;
\pi(\bb_{I^c})' \;\stackrel{\a'}\subset\; \pi_|(\aa_{I^c})' .\ee
The index of $(\a)$ is both independent of the interval $I$ by \cf\
covariance and independent of the \rep\ $\pi$ since all \rep s are
locally equivalent. We may therefore define $Index(\a)$ as the ``index
of the subtheory'' $\sub \aa\bb$.

Since the index of a subfactor equals the index of its commutant,
$(\a)$ and $(\a')$ have the same index. The index of $(\b)$ is given by
the statistical \dmn\ of $\pi$, $d(\pi)^2$, and the index of the total
inclusion $(\a'\circ\b\circ\a)$ is given by the statistical \dmn\ of the
reducible \rep\ $\pi_| \simeq \bigoplus_s N_s \pi_s$ of $\aa$,
$d(\pi_|)^2$. One can solve for the index of $\sub\aa\bb$:
\be Index[\sub\aa\bb] = \frac {d(\pi_|)}{d(\pi)} =
\frac{\sum_s N_s d(\pi_s)}{d(\pi)}. \ee
Clearly, finite index requires finite reducibility of $\pi_|$.

In coset models, the branching rules for the decomposition of a \rep\ of
$\bb$ upon restriction to $\aa$ are well known, and the statistical \dmn
s of the involved \rep s can be computed from the chiral partition
functions $\chi_\pi(\b) = Tr_\pi\exp(-2\pi\b L_0)$ as the ``asymptotic
\dmn s'' $d_{\rm as}(\pi) := \lim \chi_\pi(\b)/\chi_0(\b)$ in the
high-temperature limit $\b \searrow 0$. (Actually, the coincidence of
the asymptotic and statistical \dmn s has been established only in a few
special cases (e.g., see \cite{Wa}), but is widely believed to hold in
general. Note that the independence of (5.5) of the \rep\ $\pi$ is
manifest with the asymptotic dimension.)

The following table lists a few examples with the branching rules for
the \vrep\ (the sectors of \ca s are denoted by the corresponding
isospin, while the sectors of $W = Vir(c)$ are given in the minimal
model nomenclature) and the evaluation of (5.5):
\be \barr{ll} SU(2)_4 \subset SU(3)_1: & [0] \arr [0] + [2] \\ & Index =
1+1 = 2 \\ SU(2)_{10} \subset SO(5)_1: & [0] \arr [0] + [3] \\ & Index =
1 + \sin\frac{7\pi}{12}/\sin\frac{\pi}{12} = 3 + \sqrt 3 \\ SU(2)_2
\otimes Vir(\frac12) \subset SU(2)_1 \otimes SU(2)_1: & [0] \otimes [0]
\arr [0] \otimes [(1,1)] + [1] \otimes [(1,3)] \\ & Index = 1 \cdot 1 +
1 \cdot 1 = 2 \\ SU(2)_3 \otimes Vir(\frac7{10}) \subset SU(2)_2 \otimes
SU(2)_1: & [0] \otimes [0] \arr [0] \otimes [(1,1)] + [1] \otimes
[(1,3)] \\ & \barr{r} \!\! Index = 1 \cdot 1 + 2\cos\frac\pi 5 \cdot
2\cos\frac\pi 5 \\ = 4\cos^2\frac\pi{10}. \earr \earr \ee

As a by-product, we draw an interesting conclusion from the first and
third entries of this list. In both cases, the subtheory $\aa$ is
contained in the subtheory of $\ZZ_2$-invariants of $\bb$, where the
action of $\ZZ_2$ is the involutive Lie group \amor\ of $SU(3)$
associated with the symmetric space $SU(3)/SU(2)$, lifted to the \ca,
and the ``flip'' of the two $SU(2)_1$ tensor factors, respectively. Now,
since the index of $\sub \aa\bb$ is 2, $\aa$ must actually coincide with
the invariant subtheory. Thus, these two examples fall into the class
described by a global gauge group. One would not have expected this
result in terms of the respective Kac-Moody modes $j_n^a$. Another
example of this type, although of infinite index, is the \emt\ theory
$Vir(c=1)$ contained in the $SU(2)_1$ \ca, which in fact coincides with
the subtheory of invariants under the global $SU(2)$ \sy\ \cite{Vir}.

The last entry in the list (5.6) is an inclusion with index of the
``rigid'' Jones form $4\cos^2\frac\pi n$.

\sectreset{Discussion and outlook}
We have seen that some non-trivial sectors of a subtheory $H_l \otimes
W$ and {\it a forteriori} of $H_l$ alone are in fact generated by local
fields of the theory $G_k$. How would one have guessed the extension
$G_k$, if the physical \obs s are chosen to be a given $H_l$ theory? In
particular, what is the role of the coset theory which completely
decouples from the given $H_l$ theory? Let us begin with a simpler
question: Why have models like the last two entries in our list (5.6)
escaped the systematic search in \cite{Bul} for local extensions of
$SU(2)$ \ca s?

The answer is of course that the currents of the extending $G_k$ theory
are not primary with respect to the \emt\ $T^{(H)}$ of the original
theory, but only with respect to the \emt\ $T^{(G)}$ which according to
(1.10) is obtained by including the decoupling coset degrees of freedom.
The proper \spt\ covariance is only implemented by the latter.

For the general problem of the existence of local charged fields
generating sectors of the \obs s $H_l$, however, an extension
\be H_l \subset G_k \ee
rather than (1.9) is perfectly admissible. This incites us to take a
glimpse at infinite index. Namely, the subtheory inclusion (6.1) is
reducible with infinite index whenever $W$ in (1.9) is non-trivial,
i.e.\ $T^{\rm coset} \neq 0$. The branching of the \vrep\ comes
with infinite multiplicities, clearly falsifying Prop.\ 3. One also
readily checks in models that Prop.\ 4 and the phase condition (3.10)
are violated. This is possible since the the argument leading to (3.6)
fails. The infinite-\dmn al multiplicity space $\hh_s$ carries an {\it
infinite}-\dmn al ``internal'' \rep\ of the \mb\ by unitaries in $W$
which modifies (3.6) without affecting the intertwining property of
$\psi$ for the sectors of the observables $H_l$. This fact is the
abstract version of the coset Sugawara formula (1.10). The internal
part of the covariant transformation law for $\psi$ accounts for
$\kappa(\rho_s) \neq 1$ and for the violation of (3.10).

Thus, the reason for the introduction of the coset $W$-algebra is the
demand (principle?) of {\it local} and {\it covariant} charged fields
in the sector-generating algebra $\bb$. The latter then automatically
also contains the decoupled coset algebra.

We suggest that this sort of ``principle'' could be an alternative to
the search for a generalized gauge principle which imposes some linear
transformation law and braided \comrel s on definitely non-local and
therefore {\it a priori} un\obs\ objects. In contrast, our approach
attempts to find {\it local}\ fields which are by definition un\obs\
($\notin \aa$) with the measuring apparatus we have equipped our
observer with, but which admit the option of becoming \obs\ with an
upgraded apparatus (``breaking the \sy'' by switching on ``magnetic
fields'').

Actually, there {\it is} a way to observe the decoupled coset degrees of
freedom. Since gravity couples to the total energy-momentum density, an
observer in the $H_l$ world could detect the excess $T^{\rm coset}$ of
the total \emt\ $T^{(G)}$ beyond the \obs\ \emt\ $T^{(H)}$ through a
gravitational effect (``dark matter''). Clearly, this argument is
highly speculative. It seems not to apply in four \dmn s, where it is
known that all sectors of the algebra of \obs s are generated (actually
implemented) by charged fields (e.g.\ fermions) \cite{DR2} which are
accounted for in the standard \emt. Yet, ignoring this objection, one
could pursue a dark matter scenario like the above. It would entail that
gravitational effects belong to a different concept of observability
than the one given in terms of local quantum measurements. Thus quantum
gravity would have to stand on a very different footing than ordinary
\qft.

We observe that infinite index inclusions like (6.1) can give rise to
``anyonization'' of \bgs. Namely, a given level $k$ theory is always
diagonally embedded into the tensor product of $k$ level 1 theories
(cf.\ (1.8)), and all sectors of the former are obtained by restriction
of the sectors of the latter. But all sectors of level 1 theories for
Lie groups of Dynkin type $A,D,E$ are simple (anyonic) sectors. They
are described by (explicitly known) localized \amor s and can be
implemented by unitary operators which satisfy anyonic \comrel s.
Thus, these are models where one can further extend the local field
algebra $\bb$ into an anyonic field algebra $\ff$ such that all sectors
of $\aa$ arise in the \vrep\ of $\ff$ and the corresponding
intertwining isometries in $\ff$ satisfy anyonic \comrel s.

Finally, we include a remark answering a question raised at this
conference by S.\ Majid.
For the mathematical theory of subfactors, there is a duality between
the inclusions $\sub AB$ and $\g(B) \subset A$ and between the
associated ``paragroups'' \cite{Oc}. In particular, the same structure
theory applies for $\rho = \g\vert_A$ which is the canonical \emor\ for
the latter inclusion and $\g$ which is equivalent to the restriction of
$\rho$ to $\g(B)$. However, when extended to nets of local algebras
$\sub \aa\bb$ this balance will fail: namely in order to guarantee
$\rho$ to be a localized \emor\ of the subtheory we had to require the
compatibility of the \cex\ with the space-time symmetries $\a_x$. Now,
on one hand this implies the existence of \st, and therefore the
semigroup generated by the subsectors of $\rho$ is abelian. On the other
hand, the above compatibility is not preserved by duality: descending
the Jones tower, the \cex\ $\nu: A \arr \g(B)$ is given by $\nu(a) =
\g(V^*aV) = \lambda\mu(VV^*aVV^*)$. So the semigroup generated by $\g_t$
need not be commutative, and it definitely isn't in the case of a
non-abelian symmetry group. If one wants to maintain the mathematical
balance between the ``\sy'' paragroup (generated by $\g_t$) and the
``\sps'' paragroup (generated by $\rho_s$) in physics, either one has to
abandon localization, or the \sy\ has to be commutative (in the weak
sense of composition of equivalence classes). Is this a hint at some
unknown ``principle'' which could explain the phenomenological fact that
all exact \sps\ rules observed in Nature are due to abelian symmetries?
\\[9mm]
{\Large \bf Appendix}\\[5mm]
We sketch how Prop.\ $2'$ follows from Prop.\ 2 by entirely group
theoretical arguments.

First, the decomposition of $\g$ is just (2.9). In order to determine
the decomposition of $\rho$, we must compute the commutant $\rho\comm
A$. The map $g \mapsto |G| \cdot \mu(V\a_g(V^*))$ maps $G$ into the
commutant $\rho\comm A$:
\[ \barr{r} \mu(V\a_g(V^*)) \cdot \rho(a) = \mu(V\a_g(V^*)\rho(a)) =
 \mu(V\a_g(V^*\rho(a))) = \mu(V\a_g(aV^*)) = \quad
 \\ = \mu(Va\a_g(V^*)) = \mu(\rho(a)V\a_g(V^*)) = \rho(a)
 \cdot \mu(V\a_g(V^*)).\earr \]
By invariance of $\mu$: $\mu \circ \a_g = \mu$, and (2.2($i$)), this map
extends linearly to a $*$-\hmor\ from the group algebra $\CC G$ into
$\rho\comm A$, which is actually an \imor. Now, $\CC G$ is a direct sum
of matrix rings corresponding to the \rep s $\tau^r$ of $G$ of \dmn\
$dim(r)$. There are therefore matrix units $X^r_{ik} \in \CC G$ which
reduce the left regular \rep\ of $G$:
\[ g \; X^r_{ik} = \sum_j \ovl{\tau^r_{ji}(g)} \cdot X^r_{jk}\, . \]
Consequently, the operators
\[ a^r_{ik} := |G| \cdot \mu(V\a_{X^r_{ik}}(V^*)) \]
are algebraically matrix units in $\rho\comm A$ and can be written
in the form
\[ a^r_{ik} = W_r^i(W_r^k)^*\]
with \on\ isometries $W_r^i \in A$ yielding $\rho_r$ and the
multiplicities as in (2.13($a$)). Next, (2.2($i$)) yields
\[ \phi_r^{i*} := V^*W_r^i = V^*a^r_{ik}W_r^k =
  \a_{X^r_{ik}}(V^*) W_r^k\]
for every $k$, implying the linear transformation law (2.15) for
$\psi \propto \phi$. By (2.15), the numerical matrix
$(\phi_r^{i*}\phi_r^j)_{ij}$ commutes with the \irrep\
$\tau^r(g)$ and therefore is a multiple of $\d_{ij}$. Also by (2.15),
the average on the left-hand-side of
\[ |G|\mu(\phi_r^i\phi_r^{j*}) = (W_r^i)^* |G|\mu(VV^*) W_r^j =
(W_r^i)^* W_r^j = \delta_{ij} \]
can be computed:
\[ \sum_{g \in G} \a_g(\phi_r^i\phi_r^{j*}) = \delta_{ij}
\frac{|G|}{dim(r)} \sum_{k=1}^{dim(r)} \phi_r^k\phi_r^{k*}.\]
This establishes the completeness relations (together with the missing
normalization of $\psi \propto \phi$), such that finally (2.12) entails
(2.16).

\small \addtolength{\baselineskip}{-1pt}

\end{document}